\documentclass[a4paper,reqno]{amsart} 
\date{\today}
\usepackage{mathrsfs}
\usepackage{amsmath}
\usepackage{amsfonts}
\usepackage{amssymb}
\usepackage{amsthm}
\usepackage{amstext}
\usepackage{enumerate}
\usepackage{url}
\newcommand{\hilbert}{\mathfrak{H}}

\newcommand{\R}{\mathbb{R}}

\newcommand{\C}{\mathbb{C}}
\newcommand{\pn}{\mathcal{P}_+(\gamma)}

\newcommand{\dcero}{\mathscr{D}_0}

\newcommand{\id}{\mathbf{1}}

\newcommand{\CD}{\mathcal{D}}

\newcommand{\form}{\mathfrak{Q}}

\newtheorem{defn}{Definition}
\newtheorem{theorem}{Theorem}
\newtheorem{lemma}{Lemma}
\newtheorem{remark}[defn]{Remark}
\newtheorem*{acknowledgement}{Acknowledgement}

\begin{document}
\title{Perturbative Implementation of the Furry Picture}
 \author[M. Huber]{Matthias Huber}
\author[E. Stockmeyer]{Edgardo Stockmeyer}

\address{Mathematisches Institut\\ Ludwig-Maximilians-Universit\"at
  M\"unchen\\ Theresienstra\ss e 39\\
  80333 M\"unchen\\
  Germany}
 \email{mhuber@math.lmu.de \textrm{and} stock@math.lmu.de}
\date{July 5th, 2006}
\thanks{\copyright\ 2006 by the
       authors. This article may be reproduced in its entirety for
       non-commercial purposes.}
\begin{abstract}
  Recently 
  the block-diagonalization of Dirac-operators was investigated  
  from a mathematical point of view
  in the one-particle case \cite{SiedentopStockmeyer2005}.  We extend this result to the $N$-particle
  case. This leads to a perturbative realization of the Furry picture
  in the $N$-particle two-spinor space.\\

\end{abstract}

\maketitle
\section{Introduction}
The idea of block-diagonalizing the Dirac operator, i.e., decoupling
electronic and positronic states in such a way that the upper
components of a 4-spinor correspond to electronic and the lower
components to positronic states, goes back to Foldy and Wouthuysen
\cite{FoldyWouthuysen1950}. They succeeded in decoupling the free Dirac
operator and also addressed the case with interaction in the
non-relativistic limit. Unfortunately, their expansion does not
converge (see Thaller \cite[chapter 6]{Thaller1992} and references therein). 

The reason is that the correct parameter for the expansion, in order
to have convergence of the spectrum, is not the inverse velocity of light but
the coupling constant of the external potential. A perturbative and
iterative method to accomplish this is due to Douglas and Kroll
\cite{DouglasKroll1974} and was corrected by Jansen and He{\ss}
\cite{JansenHess1989}.  The method is very attractive for numerical
calculations because the resulting Hamiltonians operate on
two-spinors and has been successfully used for calculations in
relativistic quantum chemistry in the last twenty years 
(see \cite{BaryszSadlej2001,JansenHess1989,Kutzelnigg1997,Wolfetal2002,ReiherWolf2004a,ReiherWolf2004b}
and references therein).

From a mathematical point of view, the one-particle case was
investigated recently by Siedentop and Stockmeyer
\cite{SiedentopStockmeyer2006} (see also
\cite{SiedentopStockmeyer2005}).  They proved, under suitable
conditions on the potential, that there exists a family $U_\gamma $ of
unitary operators, analytic in the coupling constant $\gamma$, that
exactly decouples electronic and positronic states. Moreover, it was
shown that the block-diagonalized Dirac operator developed in power
series in $\gamma$ coincides, at least formally in the first orders,
with the operators resulting from the method of Douglas, Kroll and
He{\ss}. Moreover, they proved that the spectra of the truncated
expansions converge to the spectra of the original operator.

The aim of this work is to extend their results to the $N$-particle
Coulomb-Dirac Hamiltonian. We consider the $N$-particle Coulomb-Dirac
operator in the Furry picture, i.e., the operator restricted to the
positive spectral subspaces of each one particle operator. Using
simple generalizations of the methods used in
\cite{SiedentopStockmeyer2006} we prove norm resolvent convergence of
the operators resulting from the power expansion of the projected
Hamiltonian in powers of the coupling constant. For convenience for
the reader, we give in section \ref{appendix:a}  the main
definitions and results of \cite{SiedentopStockmeyer2006}, as far as
we need them.
\section{Definition of the problem and notation}\label{basicdef}
The N-particle Hilbert space is denoted by
$$\hilbert^{(N)}=\hilbert\otimes \cdots \otimes \hilbert,$$
where $\hilbert:=L^2(\R^3,\C^4)$.
An
extension of a closable operator $A$ on $\hilbert$ with domain $\CD$
to $\hilbert^{(N)}$ acting on the $j$-th component is written as
$$A_j:=\id\otimes\cdots\otimes \underbrace{A}_{\text{$j$-th place}}\otimes\cdots\otimes\id,$$
and the $n$-fold tensor product as
$$\mathcal{A}:=A\otimes\cdots\otimes A.$$
If $A$ is essentially
self-adjoint on $\CD$, then all of the above operators are essentially
self-adjoint on $\bigotimes_{j=1}^N \CD$ (see \cite{ReedSimon1972},
chapter VIII.10). For technical reasons, we will need the operator
$\mathscr{D}_0:= \sum_{j=1}^N |D_0|_{j}$, where $D_0$ is the free
Dirac operator.

The Coulomb-Dirac operator is given formally by
\begin{equation}\label{cd}
H_{CD}^{(N)}:=\sum_{j=1}^{N}[D_\gamma] _j+\frac{\gamma}{Z} \sum_{1\leq i<j\leq N} W_{ij},
\end{equation}
where, using units $\hbar=c=1$, $Z$ is the atomic number of the
nucleus, $\gamma=Z\alpha^2$ is the coupling constant of the
one-particle potential, and $\alpha$ is the fine structure constant.
$D_\gamma =D_0+\gamma V$ where $D_0$ is the free Dirac operator,
$V=-1/|\cdot|$ is the Coulomb interaction with the nucleus, and
$W_{ij}$ is the interaction between the particle $i$ and $j$, which is
defined by
\begin{equation}
(W_{ij}f)(x_1, \ldots, x_N):= \frac{f(x_1, \ldots, x_i,\ldots, x_j, \dots, x_N)}{|x_i-x_j|}
\end{equation}
for $f\in H^1(\R^3)^4\otimes \cdots \otimes H^1(\R^3)^4$. We set
$W_\gamma:=\frac{\gamma}{Z}\sum_{1\le i<j\le N} W_{ij}$.  It is well
known that the operator in \eqref{cd} (without the electron-electron
interaction $W_\gamma$) has the whole real axis as its spectrum.

We consider instead of \eqref{cd} the Coulomb-Dirac operator in the Furry picture, that is we
restrict \eqref{cd} onto the positive spectral subspaces of each
one-particle operator $D_\gamma $. The Hilbert space is given by
$$\hilbert^{(N)}_+(\gamma)=P_+^\gamma \hilbert\otimes \cdots \otimes P_+^\gamma \hilbert,$$
where $P_+^\gamma :=\chi_{[0,\infty)}(D_\gamma )$. The $N$-particle projection is given by
\begin{equation}
  \mathcal{P_+^\gamma}:=P_+^\gamma \otimes\cdots\otimes P_+^\gamma .
\end{equation}
We are interested in
$\mathcal{P_+^\gamma}H_{CD}^{(N)}\mathcal{P_+^\gamma}$ in particular,
in its realization in the $N$ particle two-spinor space. We recall the
definition and properties of the unitary transformation $U_\gamma $
given in \cite{SiedentopStockmeyer2006} (see equation \eqref{uni}
below) and consider its $N$-particle version
\begin{equation}
  \mathcal{U}_\gamma:=U_\gamma \otimes\cdots\otimes U_\gamma .
\end{equation}
This operator has the property $\mathcal{U}_\gamma\mathcal{P_+^\gamma}\mathcal{U}^{-1}_\gamma=
\mathcal{P}_{+}^{0}$. Moreover, the $N$-particle Foldy-Wouthuysen transformation
$\mathcal{U}_{\rm{FW}}$ fulfills $\mathcal{U}_{\rm{FW}}\mathcal{P}_+^0=\beta_+\otimes\cdots\otimes\beta_+ \mathcal{U}_{\rm{FW}}$
where $\beta_+:=(1+\beta)/2$ is the projection onto the upper two-spinor.
Therefore, the formal Hamiltonian  in the Furry picture realized in the $N$-particle 
two-spinor space is given by 
\begin{equation}\label{ha1}
H^N_{\rm{diag}}=\mathcal{U}_{\rm{FW}}\mathcal{U}_\gamma\mathcal{P_+^\gamma}
H_{CD}^{(N)}
\mathcal{P_+^\gamma}\mathcal{U}^{-1}_\gamma\mathcal{U}_{\rm{FW}}^{-1}.
\end{equation}
Our main result, Theorem \ref{Thm:NormResConv} below, is that
the spectra of the operators resulting from the expansion in $\gamma$
of \eqref{ha1} converge to the spectrum of the Furry-operator given formally 
by $\mathcal{P_+^\gamma}H_{CD}^{(N)}\mathcal{P_+^\gamma}$.
\section{Main result}  
The one-particle Dirac operator with Coulomb potential is given by
\begin{equation}
D_\gamma :=\boldsymbol{\alpha}\cdot
\tfrac1i\nabla+\beta-\gamma \frac{1}{|\cdot|}
\end{equation}
acting in the Hilbert space $L^2(\R)^4$. For $|\gamma|<\sqrt{3}/2$ the
operator $D_\gamma $ is essentially self-adjoint on
$C_0^\infty(\R^3\setminus\{0\})^4$ and self-adjoint on $H^1(\R^3)^4$.
Furthermore, for $\gamma<1$ the operator has a distinguished
self-adjoint extension characterized by $\mathfrak{D}(D_\gamma)\subset
H^{1/2}$ (see Thaller \cite[Theorem 4.4.]{Thaller1992}). Moreover, for
$|\gamma|<1$ we have that $|D_\gamma |\ge\sqrt{1-\gamma^2}>0$.
 
We define the operator $\tilde{T}_{\gamma}: \mathfrak{H}_+^{(N)}\to
\mathfrak{H}_+^{(N)}$ without electron-electron interaction as
\begin{equation}
\tilde{T}_{\gamma}=\mathcal{P_+^\gamma}\sum_{i=1}^N [D_\gamma] _i\mathcal{P_+^\gamma},
\end{equation}
with domain $\mathfrak{D}(\tilde{T}_\gamma)=\mathcal{P_+^\gamma}
H^{1}(\R^3,\C^4)^{\bigotimes N}$. For
$|\gamma|<\sqrt{3}/2$ the operator $\tilde{T}_\gamma$ is essentially
self-adjoint (\cite[Theorem VIII.33]{ReedSimon1972}). We denote the
closure of $\tilde{T}_\gamma$  by $T_\gamma$ and its form domain by
$\form(T_\gamma)$.

Define the quadratic form $q(f,g):=(f,T_\gamma g)+(f,W_\gamma g)$ for
$f,g\in \form(T_\gamma)$. We have the following:
\begin{theorem}\label{s.a.1} Let $0\leq \gamma<\sqrt{3}/2$.
  There exists a unique self-adjoint operator $H_{\gamma,+} :
  \hilbert_+^{(N)}(\gamma)\to \hilbert_+^{(N)}(\gamma)$ with
  $\form(H_{\gamma,+})=\form(T_\gamma)=\mathcal{P_+^\gamma}H^{1/2}(\R^{3N}, \C^{4^N})$
  where $ \mathcal{P_+^\gamma} (H^1(\R^3)^4)^{\otimes N}$ is a form core for
  $H_{\gamma,+}$, such that for all $f,g\in \form(H_{\gamma,+})$
$$q(f,g)=(f,H_{\gamma,+}g).$$
This self-adjoint extension is the Friedrich extension of the symmetric operator $T_\gamma +\mathcal{P_+^\gamma}W_\gamma\mathcal{P_+^\gamma}$ defined on $\mathcal{P_+^\gamma} \bigotimes_{j=1}^N H^1(\R^3)^4$.
\end{theorem}
\begin{proof}
\emph{Step 1: Well-definedness of $q$.} We have by Lemma  \ref{Lm:Ineq}
 \begin{equation}\label{bound1}
(f,|D_0|_{i}f)\le \frac{1}{d_\gamma}(f,|D_\gamma |_{i}f)\le \frac{1}{d_\gamma}(f,T_\gamma f).
\end{equation}
This inequality  together with Lemma \ref{Lm:IntBound} implies the well-definedness of $q$.\\
\emph{Step 2: Definition of $H_{\gamma,+}$.} We mimick the proof of the KLMN-theorem, using the notation of 
 of Reed-Simon \cite[Theorem X.17]{ReedSimon1975}. Pick $f\in \form(T_\gamma)$. 
We start by proving that $(f, W_\gamma f)\le c(f, T_\gamma f)$ for some $c>0$. Since
for $f\in\hilbert_+^{(N)}$, we have
\begin{equation}\label{y1}
\begin{split}
(T_\gamma^{-1/2}f,& W_\gamma T_\gamma^{-1/2}f)=\\
&=\frac{\gamma}{Z}\sum_{1\le i<j\le N} 
(T_\gamma^{-1/2}f,|D_0|_i^{1/2} |D_0|_i^{-1/2}W_{ij}|D_0|_i^{-1/2}|D_0|_i^{1/2}
 T_\gamma^{-1/2}f)\\
 &\le \frac{\gamma\pi}{2Z}\sum_{1\le i<j\le N} (T_\gamma^{-1/2}f, |D_0|_i T_\gamma^{-1/2}f)
\le \frac{\gamma\pi N(N-1)}{4Zd_\gamma}\|f\|^2=:c\|f\|^2,
\end{split}
\end{equation}
where we used Lemma \ref{Lm:IntBound} and equation \eqref{bound1}. Therefore,
\begin{equation} \label{Ineq:FormEquiv1}(f,T_\gamma f)+(f,W_\gamma f)\le (1+c)(f,T_\gamma f)\le
(1+c)(f,(T_\gamma+W_\gamma)f).\end{equation}
The latter shows that the norms
$\|\cdot\|_{+1,T_\gamma}$ and $\|\cdot\|_{+1,q}$ are equivalent. Thus $q$ is
a semi-bounded, closed quadratic form on $\form(T_\gamma)$, which 
therefore defines a self-adjoint operator $H_{\gamma,+}$ with 
form-domain $\form(H_{\gamma,+})=\form(T_\gamma)$.\\
\emph{Step 3: Determination of the form domain.}
We have the following chain of inequatlities, where we used Lemma \ref{Lm:Ineq} in the second and Hardy's inequality in the third step:
\begin{multline}\label{Ineq:FormEquiv}
  d_\gamma \sqrt{-\triangle_{3N}+1}\leq d_\gamma \sum_{i=1}^N |D_0|_i
  \leq \sum_{i=1}^N |D_\gamma|_i\leq\\ \sum_{i=1}^N (1+2\gamma )
  |D_0|_i\leq (1+2\gamma ) N \sqrt{-\triangle_{3N}+1}
\end{multline}
Inequalities \eqref{Ineq:FormEquiv} imply that $H^{1/2}(\R^{3N},
\C^{4^N})$ is complete with respect to the quadratic form
$b(f,f):=(\sqrt{\sum_{i=1}^N |D_\gamma|_i}f, \sqrt{\sum_{i=1}^N
  |D_\gamma|_i}f)$ for $f\in H^{1/2}(\R^{3N}, \C^{4^N})$. Since
$\mathcal{P_+^\gamma}$ commutes with $\sum_{i=1}^N |D_\gamma|_i$, we
have $\mathcal{P_+^\gamma}H^{1/2}(\R^{3N}, \C^{4^N})\subset
H^{1/2}(\R^{3N}, \C^{4^N})$. Moreover,
$\mathcal{P_+^\gamma}H^{1/2}(\R^{3N}, \C^{4^N})$ is a closed subspace
of $H^{1/2}(\R^{3N}, \C^{4^N})$ with respect to the norm generated by
$b$. Hence $\mathcal{P_+^\gamma}H^{1/2}(\R^{3N}, \C^{4^N})$  is complete 
with respect to the restriction of $b$. Since $\tilde{T}_\gamma$ is essentialy self-adjoint
the self-adjoint operator associated to the restriction of $b$
is $T_\gamma$.  Thus, $\form(T_\gamma)=\mathcal{P_+^\gamma}H^{1/2}(\R^{3N}, \C^{4^N})$. \\

Since obviously
$\mathfrak{D}(H^{N}_{\gamma, +})\subset \form(H^{N}_{\gamma,
  +})=\mathcal{P_+^\gamma}H^{1/2}(\R^{3N}, \C^{4^N})$, we have
that $H^{N}_{\gamma, +}$ is the Friedrich extension of the symmetric
operator $\tilde{T}_\gamma
+\mathcal{P_+^\gamma}W_\gamma\mathcal{P_+^\gamma}$ defined on $\mathcal{P_+^\gamma}
\bigotimes_{j=1}^N H^1(\R^3)^4$.
\end{proof}
We now turn to the operator $\mathcal{U}_\gamma:=U_\gamma
\otimes\cdots\otimes U_\gamma $ which is a unitary mapping
$\mathcal{U}_\gamma : \hilbert_+^{(N)}(\gamma) \to
\hilbert_+^{(N)}(0)$. We define the operator $\tilde{H}_{\gamma,+} :
\hilbert_{+,0}^{(N)}\to \hilbert_{+,0}^{(N)}$ as $\tilde{H}_{\gamma,+}
:=\mathcal{U}_\gamma H_{\gamma,+}\mathcal{U}_\gamma^{-1}$, where
$\hilbert_{+,0}^{(N)}:=\hilbert_+^{(N)}(0)$.  Analogously, 
interpreting the Foldy-Wouthuysen transformation in a natural way as a
mapping $\mathcal{U}_{\rm{FW}}: \hilbert_{+,0}^{(N)}\to
L^2(\R^3,\C^2)^{\otimes N}=L^2(\R^{3N},\C^{2^N})$, we define
$H_{\gamma,+}^{\rm{diag}}: L^2(\R^{3N},\C^{2^N})\to L^2(\R^{3N},\C^{2^N})$ as the operator
$H_{\gamma,+}^{\rm{diag}}=\mathcal{U}_{\rm{FW}}
\tilde{H}_{\gamma,+}\mathcal{U}_{\rm{FW}}^{-1}$. In this way,
$H_{\gamma,+}^{\rm{diag}}$ can be seen as the block-diagonalization of
$\tilde{H}_{\gamma,+}$.  By Theorem \ref{s.a.1},
$\tilde{H}_{\gamma,+}$ and $H_{\gamma,+}^{\rm{diag}}$ are self-adjoint
operators with form domain $\mathcal{U}_\gamma\form(T_\gamma)$ and
$\mathcal{U}_{\rm{FW}}\mathcal{U}_\gamma\form(T_\gamma)$ respectively.  Actually $\mathcal{U}_\gamma\form(T_\gamma)=\mathcal{P}^0_+H^{1/2}(\R^{3N}, \C^{4^N})$, since
$\dcero^{1/2}\mathcal{U}_\gamma\dcero^{-1/2}$ is a bounded operator
(see proof of Lemma \ref{new}). Since $\mathcal{U}_{\rm{FW}}$ commutes with $\mathscr{D}_0$, we get
$\mathcal{U}_{\rm{FW}}\mathcal{U}_\gamma\form(T_\gamma)=H^{1/2}(\R^{3N}, \C^{2^N})$.

We denote 
by $\tilde{h}_{\gamma,+}^{k}$ the 
formal Taylor expansion of $H_{\gamma,+}^{\rm{diag}}$ up to the power 
$\gamma^k$ inclusive, acting on  
$\mathcal{C}:=\mathcal{U}_{\rm{FW}}\mathcal{U}_\gamma\pn
H^1(\R^3,\C^4)^{\otimes N}$ which is  a form core for $H_{\gamma,+}^{\rm{diag}}$.
We set $R_\gamma^k=H_{\gamma, +}^{\rm{diag}}-\tilde{h}_{\gamma,+}^{k}$.

 The main result of this letter is:
\begin{theorem}\label{Thm:NormResConv}
  There exists a $\gamma_c>0$ such that for $0\leq\gamma < \gamma_c$
  the operators $\tilde{h}_{\gamma,+}^{k}$ admit a distinguished
  self-adjoint extension $h_{\gamma,+}^{k}$ for $k$ big enough with
  the property that $\mathfrak{D}(h_{\gamma,+}^{k})\subset
  \mathfrak{Q}(H_{\gamma,+}^{\rm{diag}})=H^{1/2}(\R^{3N}, \C^{2^N})$. Moreover
  $h_{\gamma,+}^{k}\rightarrow H_{\gamma,+}^{\rm{diag}}$ as $k\to\infty$ in the sense of norm resolvent
  convergence.
\end{theorem}

\begin{proof}
 According to Kato \cite[Theorem VI.3.11 and Corollary VI.3.12]{kato1966} 
 it is enough to prove there exist a sequence $a_k$ with $a_k\to 0$ as $k\to \infty$, such that for any
$f\in\mathcal{C}$
\begin{equation}\label{q1}
(f,R_\gamma^k f )\le a_k(f,H_{\gamma, +}^{\rm{diag}}f).
        \end{equation}
This is equivalent to 
\begin{equation}\label{q2}
(f,H_{\gamma,+}^{-1/2}\mathcal{U}^{-1}_\gamma\mathcal{U}_{\rm{FW}}^{-1}R_\gamma^k  \mathcal{U}_{\rm{FW}}\mathcal{U}_\gamma H_{\gamma,+}^{-1/2}f)\to 0
\end{equation}
for $f\in \pn H^1(\R^3, \C^4)^{\otimes N}$.
 Then
\begin{equation}\label{q3}
\begin{split}
(f,&H_{\gamma,+}^{-1/2}\mathcal{U}^{-1}_\gamma\mathcal{U}_{\rm{FW}}^{-1}R_\gamma^k  \mathcal{U}_{\rm{FW}}\mathcal{U}_\gamma H_{\gamma,+}^{-1/2}f)\\
&\le \|\mathscr{D}_0^{-1/2}R_\gamma^k \mathscr{D}_0^{-1/2}\|\,
\|\mathscr{D}_0^{1/2}\mathcal{U}_{\gamma}\mathscr{D}_0^{-1/2}\|^2\,
\|\mathscr{D}_0^{1/2}H_{\gamma,+}^{-1/2}f\|^2,  
\end{split}
\end{equation}
the last term goes to zero due to lemmas \ref{Thm:Analyticity}, \ref{new} and \ref{new2} below. We used in the last step that $U_{FW}$ commutes with $|D_0|$ and therefore  $\mathcal{U}_{FW}$ with $\mathscr{D}_0$ and any function of it.
\end{proof}
\begin{remark}
The unitary transform $U_\gamma $ is not unique. If we take the choice
given in \cite{SiedentopStockmeyer2006} \eqref{uni} in the appendix, we know that $\gamma_c\ge 0,3775$
which corresponds to the critical atomic number $Z=52$.
\end{remark}
\section{Auxiliary lemmas}
The following bound
is Kato's inequality.
\begin{lemma}\label{Lm:IntBound}
    Let $f\in  H^{1/2}(\R^3N, \C^{4^N})$. Then
    \begin{equation}
        (f,W_{ij}f)\leq \frac{\pi}{2}(f,|D_{0}|_{j}f) \text{ for $i\neq j$}
    \end{equation}
\end{lemma}
\begin{proof}
    Pick  $f\in \bigotimes_{j=1}^N H^1(\R^3)^4 $ arbitrarily.
    \begin{multline}
        (f,W_{ij}f)=\int dx_1 \cdots dx_N \overline{f(x_1,\ldots, x_i,\ldots, x_N)}\frac{1}{|x_i-x_j|}
        f(x_1,\ldots, x_i,\ldots, x_N)=\\
        \int dx_1 \cdots dx_N \overline{f(x_1,\ldots, x_i+x_j,\ldots,  x_N)}
        \frac{1}{|x_i|}f(x_1,\ldots, x_i+x_j,\ldots, x_N)\leq\\\frac{\pi}{2}
        \int dx_1 \cdots dx_N \overline{f(x_1,\ldots, x_i+x_j,\ldots,  x_N)}
        |D_{0}|_{i} f(x_1,\ldots, x_i+x_j,\ldots, x_N)\leq\\
        \frac{\pi}{2}(f,|D_{0}|_{i}f),
\end{multline}
where we used Kato's inequality. This inequality extends by continuity to $f\in H^{1/2}(\R^3N, \C^{4^N})$.
\end{proof}

\begin{lemma}\label{Thm:Analyticity} For $ |\gamma|<\gamma_c$ 
  the operator  $\mathscr{D}_0^{-1/2}{H}_{\gamma,+}^{\rm{diag}}
  \mathscr{D}_0^{-1/2}$ is bounded and real analytic around zero. In particular
  $\|\mathscr{D}_0^{-1/2}R_\gamma^k\mathscr{D}_0^{-1/2}\|\to
  0$.
\end{lemma}
\begin{proof} First note that $|D_0|_i^{1/2}\mathscr{D}_0^{-1/2}$
  is bounded. To prove this, we take an arbitrary $f\in
  \bigotimes_{j=1}^N H^1(\R^3)^4$ and note that
\begin{multline}
\||D_0|_i^{1/2} f\|^2=(f, |D_0|_{1} f)\leq (f, |D_0|_{1} f) +\cdots+ (f,
|D_0|_{N} f)=\|\dcero^{1/2}f\|^2,
\end{multline}
the claim follows by density of $\bigotimes_{j=1}^N H^1(\R^3)^4$ in $\hilbert^{(N)}$. 

According to Lemma \ref{Lm:UAnal1}, the operator $|D_0|^{-1/2}U_\gamma
D_\gamma U_\gamma ^*|D_0|^{-1/2}$ is bounded and analytic in $\gamma$
for $|\gamma|<\gamma_c$, and so is $[(|D_0|^{-1/2}U_\gamma D_\gamma
U_\gamma ^*|D_0|^{-1/2})]_m $ for $m=1,\ldots, N$. In order to prove
the claim
it suffices to show the analyticity of each summand of $\tilde{H}_{\gamma,+}$.\\
\emph{One-particle terms:} The analyticity follows immediately from
\begin{multline}
  \dcero ^{-1/2}[ U_\gamma P_+^\gamma D_\gamma P_+^\gamma
  U^{-1}_\gamma]_m \dcero ^{-1/2}= \mathcal{P}_+^0\dcero^{-1/2}
  [|D_0|^{1/2} ]_{m}\\ \times[|D_0|^{-1/2}U_\gamma D_\gamma U_\gamma
  ^*|D_0|^{-1/2}]_m |D_0|^{1/2}_{m} \dcero^{-1/2}\mathcal{P}_+^0.
\end{multline}
\emph{Interaction terms:}
The operator 
$$|D_0|^{-1/2}_{m}\mathcal{U}_\gamma|D_0|^{1/2}_{m}= U_\gamma
\otimes\cdots\otimes |D_0|^{-1/2} U_\gamma |D_0|^{1/2} \cdots\otimes
U_\gamma$$
is analytic by Lemma \ref{Lm:UAnal1}, and
$|D_0|^{-1/2}_{m}W_{m,l} |D_0|^{-1/2}_{m}$ is bounded by Lemma
\ref{Lm:IntBound}.  Thus, writing
\begin{multline}
\dcero^{-1/2} \mathcal{U}_\gamma\mathcal{P_+^\gamma}W_{m,l}\mathcal{P_+^\gamma} \mathcal{U}^{-1}_\gamma
 \dcero^{-1/2}=
\mathcal{P}_+^0 \dcero^{-1/2} \mathcal{U}_\gamma W_{m,l} \mathcal{U}^{-1}_\gamma
 \dcero^{-1/2}\mathcal{P}_+^0=\\
\mathcal{P}_+^0\dcero^{-1/2} |D_0|^{1/2}_{m}
 |D_0|^{-1/2}_{m}\mathcal{U}_\gamma|D_0|^{1/2}_{m}
|D_0|^{-1/2}_{m} W_{m,l} |D_0|^{-1/2}_{m}\\
|D_0|^{1/2}_{m} \mathcal{U}^{-1}_\gamma|D_0|^{-1/2}_{m}
 |D_0|^{1/2}_{m}   \dcero^{-1/2}\mathcal{P}_+^0,
\end{multline}
 we have shown the analyticity of the interaction term.
 
 Since $\mathcal{U}_{\rm{FW}}$ commutes with $\mathscr{D}_0$ the operator $\mathscr{D}_0^{-1/2}H_{\gamma,+}^{\rm diag}\mathscr{D}_0^{-1/2}$
 is also analytic and therefore has a convergent Taylor expansion with $\|\mathscr{D}_0^{-1/2}R_\gamma^k\mathscr{D}_0^{-1/2}\|\to
  0$ as $k\to\infty$.
\end{proof}

\begin{lemma}\label{new}
  For $|\gamma|<\gamma_c$ the operator
  $\dcero^{1/2}\mathcal{U}_\gamma\dcero^{-1/2}$ is bounded on
  $\hilbert^{(N)}_0$.
\end{lemma}
\begin{proof}
  We have that $\displaystyle \dcero^{1/2} \le \sum_{i=1}^N
  |D_0|_i^{1/2}.$ Note further that by Lemma \ref{Lm:UAnal1} the
  operator $|D_0|^{1/2}U_\gamma |D_0|^{-1/2}$ is bounded, thus
\begin{equation}\begin{split}
    \|\dcero^{1/2}\mathcal{U}_\gamma\dcero^{-1/2}\|&=\|\dcero^{1/2}(\sum_{i=1}^N |D_0|_i^{1/2})^{-1}\sum_{i=1}^N |D_0|_i^{1/2}\mathcal{U}_\gamma\dcero^{-1/2}\|\\
    &=\|\dcero^{1/2}(\sum_{i=1}^N |D_0|_i^{1/2})^{-1}\sum_{i=1}^N (|D_0|_i^{1/2}\mathcal{U}_\gamma|D_0|_i^{-1/2}|D_0|_i^{1/2}\dcero^{-1/2})\|\\
    &\le\sum_{i=1}^N \||D_0|_i^{1/2}\mathcal{U}_\gamma|D_0|_i^{-1/2}|D_0|_i^{1/2}\dcero^{-1/2})\|\\
    &\le\sum_{i=1}^N
    \||D_0|_i^{1/2}\mathcal{U}_\gamma|D_0|_i^{-1/2}\|\||D_0|_i^{1/2}\dcero^{-1/2}\|
\end{split}
\end{equation}
is finite.
\end{proof}
\begin{lemma}\label{new2} For $f\in \mathfrak{H}^{(N)}_+(\gamma)$, and $0\le\gamma<\sqrt{3}/2$ we 
  have the estimate $\|\dcero^{1/2} H_{\gamma,+}^{-1/2}f\|\le
  1/\sqrt{d_\gamma} \|f\|$.
\end{lemma}
\begin{proof}
  Pick $f\in \mathcal{P}_+(\gamma)\left(H^1(\R^3)^4\right)^{\otimes
    N}$. Then using Lemma \ref{Lm:Ineq} and $W_\gamma\ge 0$ we get
\begin{equation}
(f, \dcero f)\leq \frac{1}{d_\gamma}(f, \sum_j (P_+^\gamma D_\gamma  P_+^\gamma)_j f)\leq  \frac{1}{d_\gamma}(f, H_{\gamma,+} f).
\end{equation} 
\end{proof}
\section{The one-particle case}\label{appendix:a}
Let us define the following operator 
\begin{equation}\label{uni}
U_\gamma :=(P_+^0 P_+^\gamma+P_-^0
P_-^\gamma)(1-(P_+^0-P_+^\gamma))^2)^{-1/2}.
\end{equation}
Some important
properties (\cite[Theorem 1, Theorem 2, Lemma
9]{SiedentopStockmeyer2006}) of $U_\gamma $ are listed in the
following lemma. We set $\gamma_c:=0.3775$.
\begin{lemma}\label{Lm:UAnal1}
\begin{enumerate}
\item $U_\gamma $ is analytic in $\gamma$ and unitary for
  $|\gamma|<0.6841$ and fulfills the relation
  $$U_\gamma P_\pm^\gamma=P_\pm^0U_\gamma .$$
\item The operator $|D_0|^{1/2}U_\gamma |D_0|^{-1/2}$ is bounded and
  analytic in $\gamma$ for $\gamma<\gamma_c$.
\item The operator $|D_0|^{-1/2}U_\gamma D_\gamma  U_\gamma ^\ast
  |D_0|^{-1/2}$ is bounded and analytic in $\gamma$ for
  $\gamma<\gamma_c$.
\end{enumerate}
\end{lemma} 
The following inequality is used in this paper.
For $0\leq\gamma< \frac{\sqrt{3}}{2}$ set $C_{\gamma}:=\frac{1}{3}\left(\sqrt{4(\gamma)^2+9}-4\gamma\right)$ and 
$d_{\gamma}:=\frac{1}{2}(1+C_{\gamma}^2-\sqrt{(1-C_{\gamma}^2)^2+4\gamma^2 C_{\gamma}^2})$. We have (see Morozov \cite{Morozov2004} and also Brummelhuis et al \cite{BRS2001} ):
\begin{lemma}\label{Lm:Ineq}
    For $0\leq\gamma< \frac{\sqrt{3}}{2}$ the operator inequality
    \begin{align}\label{DFCoulombIneqIneq1}
              |D_\gamma |^2&\geq d_{\gamma}^2|D_0|^2
    \end{align}
holds.
\end{lemma}

\begin{acknowledgement}
Both authors acknowledge 
financial support from the  EU IHP network 
{\it Postdoctoral Training Program in Mathematical Analysis of
Large Quantum Systems},
contract no.\
HPRN-CT-2002-00277. M.H. acklowledges support by the Deutsche Forschungsgemeinschaft (DFG), grant no. SI 348/12-2.
\end{acknowledgement}

\end{document}